\documentclass[aps,pre,twocolumn,showpacs,amssymb]{revtex4}

\usepackage[dvips]{graphicx}
\usepackage{dcolumn}
\usepackage{bm}

\def\gs{\gtrsim}
\def\ls{\lesssim}
\def\be{\begin{equation}}
\def\en{\end{equation}}                  
\def\p{\partial} 
\newcommand{\bi}[1]{\mbox{\boldmath$#1$}}\newcommand{\av}[1]{\langle{#1}\rangle}\def\bea{\begin{eqnarray}}
\def\ena{\end{eqnarray}}

\def\ge{> \kern -12pt \lower 5pt \hbox{$\displaystyle =$}}
\def\le{< \kern -12pt \lower 5pt \hbox{$\displaystyle =$}}
\def\gs{> \kern -12pt \lower 5pt \hbox{$\displaystyle{\sim}$}}
\def\ls{< \kern -12pt \lower 5pt \hbox{$\displaystyle{\sim}$}}

\def\ve{\varepsilon}

\begin{document}
\title{Charged inclusion in nematic liquid crystals}
\author{Lionel Foret and Akira Onuki}
\affiliation{Department of Physics, Kyoto University, Kyoto 606-8502, Japan}

\begin{abstract}
We  present a general theory of 
liquid crystals under inhomogeneous electric field 
in a Ginzburg-Landau scheme. 
The molecular orientation can be deformed by  electric field 
when the dielectric tensor is orientation-dependent. 
We then investigate the influence of a charged 
particle on  the orientation order 
 in  a nematic state.  
The director  is aligned   either 
along or perpendicular to 
the local electric field around the charge, 
depending on the sign of the dielectric anisotropy. 
The  deformation becomes stronger with increasing 
the ratio $Ze/R$,  where $Ze$ is the charge and $R$ is the radius 
of the particle. Numerical 
analysis shows  the presence  of 
defects around the particle for large $Ze/R$. 
They are nanometer-scale defects for microscopic ions.  
If the dielectric anisotropy is positive, 
  a Saturn ring defect appears. If it is negative, 
a pair of point defects appear 
apart from the particle surface,  each being 
connected to the surface by a disclination line segment.

\end{abstract}
\pacs{61.30.Dk, 61.30.Jf, 77.84.Nh, 61.30.Gd}  
\maketitle

\section{Introduction}

Recently, a number of complex mesoscopic 
structures have been observed with addition of  small 
particles  into a  liquid crystal matrix 
\cite{Pou97,Zap99}. In nematics, 
inclusions can distort the orientation order over long 
distances,  often inducing  
 topological defects 
 \cite{Lav00,Ter95,Lub98,P1,Allen,Yama}. As  a result, 
anisotropic log-range interactions 
mediated by the distortion are  produced 
among the immersed particles. 
It also leads to the formation of 
various uncommon structures or phases, 
such as chain aggregates \cite{Pou97}, soft solids 
supported by a jammed cellular network 
of particles \cite{Mee00}, 
or a  transparent phase 
including microemulsions  \cite{Yam01,Bel03}.

The short-range  anchoring of the nematic molecules on 
the inclusions surface is usually taken as the origin 
of the long-range distortion 
\cite{Lav00,Ter95,Lub98,P1,Allen,Yama,Sta04,Fuk04}. 
In the continuum approach 
in terms of the director $\bi n$  in a nematic state, 
the   anchoring  free energy   is 
expressed as the surface integral, 
\be 
F_a= -\frac{1}{2}  W_a \int dS
 ({\bi n}\cdot{{\bi \nu}})^2,
\en 
where $dS$ is the surface element, 
  ${{\bi \nu}}$ is the normal unit vector 
at the  surface, and  $W_a$ is a parameter 
representing  the strength of anchoring \cite{anchor}.  
As another anchoring  mechanism,  
electrically charged inclusions  
 should induce  alignment  of 
the nematic order in their vicinity \cite{NATO,Onu04}. 
This mechanism  is  relevant for     
  ions which are naturally present or 
externally doped  and  for colloidal particles 
whose surfaces are  highly charged. 
 A large 
charged  particle can  also be inserted 
into     nematics.    However,   the effect of charges  in  
liquid crystals remains poorly understood 
and has rarely been studied, 
despite its obvious fundamental and technological importance. 
It is of great interest how the 
charge anchoring mechanism works and how it is different 
from the usual short-range anchoring mechanism.

It is also worth noting that  the ion mobility 
in nematics is  known to be anomalously low as 
compared  to  that  in liquids with  similar 
viscosity \cite{Gen93}. Originally,  de Gennes \cite{PG}  
  attributed the origin of this observation   
 to a  long-range deformation  
of the orientation order of 
the surrounding liquid crystal  molecules. 
Analogously,  small ions such as Na$^+$ or Cl$^-$   in polar fluids    
are surrounded by a   microscopic solvation shell  composed 
of  polar molecules aligned along the local electric field 
\cite{Born,Is,Kit04}.

The coupling of the electric field and  the 
nematic orientation arises from 
the fact that the  dielectric tensor $\ve_{ij}$  
depends on  the director $\bi n$  or 
the local orientation tensor  $Q_{ij}$.  
The alignment along a homogeneous  electric field 
is a well-known effect \cite{Gen93}, while 
  the alignment in an  inhomogeneous electric field 
 poses   very complicated problems. 
We  mention an experiment \cite{P2}, in which 
an electric field was applied 
to   nematics  containing  
silicone oil particles to produce  
 field-dependent defects.  
In this paper, we will  study 
the orientation deformation 
around a single charged particle  in a nematic state 
in the phenomenological  Landau-de Gennes scheme using 
the  orientation tensor $Q_{ij}$ 
\cite{Scho,Hess}.   The  optimal  alignment 
  minimizes  the sum of the Landau-de Gennes 
free energy and the electrostatic energy. 
Similar approaches have recently been used 
to calculate  the  solvation free energy of 
ions in near-critical polar 
fluids \cite{Kit04}.

This  paper is organized as follow. 
In  Sec.II, we will present a general 
Ginzburg-Landau   framework. 
By  minimizing the free energy functional,  we will 
obtain   equilibrium equations satisfied by  $Q_{ij}$, while  
the electric potential  obeys the Poisson equation   
with a  dielectric tensor  dependent on $Q_{ij}$.
In Sec.III, we will estimate the 
free energy contributions around 
an isolated  charged particle 
to find the range of the orientation deformation 
and the condition of strong deformation. 
Some discussion will be given on the anchoring 
effect around a charged  colloidal particle 
surrounded by counterions. 
 We will also give some discussions 
for  a charged  colloidal particle surrounded by  counterions.    
  In  Sec.IV, we 
will  numerically solve the equilibrium  equations.  We shall find     
 the formation of  topological defects 
for  large charge $Ze$ and/or small radius $R$ 
of the particle.

\setcounter{equation}{0} 
\section{General theoretical background}

\subsection{Free energy functional with charges}

The liquid crystal  order is described in terms of  
the symmetric, traceless orientation 
tensor $Q_{ij}$ \cite{Gen93,Scho,Hess}, 
which may be  defined as  
\be
Q_{ij}
(\bi{r}) = \av{u_{i} u_{j} - \delta_{ij}/3 },
\label{eqI1}
\end{equation}
where $\bi{u}$ denotes the unitary 
orientation vector of the nematic molecules. 
We introduce  rotationally invariant quantities, 
\be 
J_2= \sum_{ij}Q_{ij}^2, \quad 
J_3=\sum_{ijk}Q_{ij}Q_{jk}Q_{ki}.
\en 
We assume the Landau-de Gennes  free energy $F_{\rm L}=F_0+F_g$ 
\cite{Gen93}   consisting  of two parts,  
\bea
F_{0}& =&\int d{\bi r}\bigg [-\frac{A}{2}J_2 -
\frac{B}{3}J_3 +\frac{C}{4}
J_2^2 \bigg ] ,\\
F_g&=&\int d{\bi r}
\frac{L}{2} \sum_{ijk} (\nabla_k Q_{ij}  )^2,
\ena
where 
$\nabla_k= \p /\p x_k$ with 
$x_1=x$, $x_2=y$, and $x_3=z$.  
Here $A$ is a temperature-dependent 
constant and is positive in the nematic phase, 
while $B$, $C$, and $L$ 
are positive constants in the isotropic and nematic 
phases. The transition is first-order 
for nonvanishing $B$,  and  $L\sim k_BT/\sigma$ 
with $\sigma$ being the microscopic molecular length. 
The second part $F_g$ is the gradient free energy 
representing an increase of the free energy due 
to inhomogeneity of $Q_{ij}$. We neglect another gradient 
term proportional to 
 $\int d{\bi r}\sum_{ij}|\nabla_j Q_{ij}|^2$ \cite{Gen93}.

Next we consider the free energy contribution 
$F_{e}$ arising from the  electrostatic interaction. 
It depends on the charge density $\hat{\rho}({\bi r})$ and 
the polarization vector ${\bi p}({\bi r})$ 
of the liquid crystal molecules and 
is written as \cite{NATO,Onu04}
\be 
F_{e}=\int d{\bi r}\bigg [ 
\frac{1}{2} \sum_{ij} \chi^{ij} p_ip_j + 
\frac{1}{8\pi}{\bi E}^2 \bigg ],
\en 
where ${\bi E}=-\nabla\phi$ is the electric field 
with $\phi({\bi r})$ being the electrostatic potential.  
The electric induction $\bi D= {\bi E}+ 4\pi {\bi P}$ 
satisfies 
\be 
\nabla \cdot {\bi D}= 
-\nabla^2\phi + 4\pi \nabla\cdot{\bi p}=  4\pi \hat{\rho},  
\label{eq:2.4}
\en
where $\hat{\rho}({\bi r})$ is the charge density.  
In our theory,  it is  crucial that  
the tensor $\chi^{ij}$ in Eq.(2.5) depends  on  $Q_{ij}$ 
(see Eqs.(2.11) and (2.13) below).  
However, we   assume that  it is  independent of  $\bi E$, 
 neglecting  the nonlinear dielectric effect \cite{nonlinear}. 
We assume no externally applied electric field 
and set  
\be 
\phi=0,
\en  
on the boundary walls of the container.

If infinitesimal space-dependent deviations 
$\delta {\bi p}$ and  $\delta\hat{\rho}$  
are superimposed on ${\bi p}$ and $\hat{\rho}$, 
the incremental change of  
$F_e$ is  given by 
\bea 
\delta F_e &=&
 \int d{\bi r}\bigg [
\sum_{ij}  
\bigg ( \frac{1}{2} p_ip_j\delta\chi^{ij} + 
p_j\chi^{ij}\delta p_i\bigg )\nonumber\\ 
&&\hspace{1cm}-{\bi E}\cdot{\delta {\bi p}}+\phi\delta\hat{\rho}
\bigg ], 
\label{eq:2.3}
\ena 
where we have used the relation, 
\be 
\frac{1}{2}\delta ({\bi E}^2)= 
-\nabla (\phi\delta{\bi D})+4\pi(\phi \delta\hat{\rho}-{\bi E}\cdot\delta{\bi p}).
\en  
We  minimize $F_e$ 
with respect to $\bi p$ at fixed $\hat{\rho}$ and $Q_{ij}$. 
From $\delta F_e/\delta {\bi p}={\bi 0}$ we obtain 
\be 
\sum_j\chi^{ij}p_j=E_i\quad {\rm or}\quad 
p_i= \sum_j \chi_{ij}E_j .
\en 
Thus the inverse matrix of 
$\chi^{ij}$ is 
 the electric susceptibility tensor 
$\chi_{ij}$   related to  the dielectric tensor 
${\ve}_{ij}$ by  
\be 
\ve_{ij}=\delta_{ij} +4\pi \chi_{ij}.  
\label{eq:2}
\en 
The electric induction 
is of the usual form $D_i= \sum_j \ve_{ij}E_j$ and 
the Poisson equation (2.6) becomes 
\be
\nabla\cdot{\bi D}= -\sum_{ij}\nabla_i\ve_{ij}\nabla_j\phi=4\pi\hat{\rho}. 
\label{eqI6}
\en
In this paper  the dielectric tensor is 
assumed to be linearly dependent  on $Q_{ij}$ as  
\be
 \ve_{ij} ({\bi r}) = \ve_0 \delta_{ij} +
\varepsilon_1 Q_{ij}({\bi r}), 
\en 
where $\ve_0$ is  positive,  but  
$\ve_1$ is  positive or negative depending on the 
molecular  structure  \cite{Gen93}. 
Now, from Eq.(2.10),  $F_e$ is of the standard form, 
\be 
F_{e} =  \frac{1}{8\pi}  \int d{\bi r}\sum_{ij}
{\ve_{ij}}  E_iE_j . 
\en 
We obtain  $\delta \chi^{ij}= - \sum_{k\ell}
\chi^{ik}\chi^{j\ell}\delta\chi_{k\ell}$  
from $\sum_j \chi_{ik}\chi^{kj}= \delta_{ij}$ 
in the right-hand side of 
Eq.(2.8).  Use of $\delta\chi_{ij}= \ve_1 \delta Q_{ij}/4\pi$ 
and  Eq.(2.10)  yields   
\be
\delta F_e =
 \int d{\bi r}\bigg [\phi\delta\hat{\rho}- 
\sum_{ij}  
 \frac{ \ve_1}{8\pi} E_iE_j\delta Q_{ij}  
\bigg ]. 
\en

The total free energy is the sum 
$F=F_0+F_g+ F_e$.  In equilibrium  we impose 
 the minimum condition   
$\delta F/\delta Q_{ij}= \lambda \delta_{ij}$,  
where $\lambda$ is the Lagrange multiplier 
 ensuring  the traceless condition 
$\sum_i Q_{ii}=0$.   Then Eqs.(2.3), (2.4)  and (2.15) give  
\bea 
&&(-A+CJ_2 ) Q_{ij}- 
B\bigg (\sum_k Q_{ik}Q_{kj}- \frac{1}{3}J_2\delta_{ij}\bigg )\nonumber\\ 
&&-  L\nabla^2Q_{ij}=
\frac{ \varepsilon_1}{8\pi}\bigg (E_i E_j 
-\frac{1}{3}{\bi E}^2 \delta_{ij}\bigg ), 
\ena 
where we have eliminated  $\lambda$   making   all the terms   
traceless. The field-induced  change of 
$Q_{ij}$  arises from 
the term on the right-hand side bilinear in $\bi E$. 
It is nonvanishing only for $\ve_1\neq 0$.  
In this work we are interested in the case of highly 
inhomogeneous $\bi E$  in the nematic phase. 
In the simplest  case of weak, homogeneous $\bi E$ in 
the isotropic phase, $Q_{ij}$ is  simply given by 
the right-hand side of Eq.(2.16)  divided by $|A|$  
to leading order in the field \cite{Gen93,Puz}.

\subsection{Uniaxial and biaxial orientations}

We here diagonalize the orientation tensor $\tensor Q$. 
Let the maximum eigenvalue of 
 $\tensor Q$ be written as $2S_1/3$. 
Then  the other two eigenvalues 
may be written as $-S_1/3 \pm S_2$  with $S_1\geq |S_2|$. 
Use of the principal unit vectors, 
 $\bi n$, $\bi m$, and ${\bi \ell}={\bi n}\times{\bi m}$,  
diagonalizing ${\tensor Q}$ yields  
\be 
{\tensor Q}= S_1( {\bi n}{\bi n}-\frac{1}{3}{\tensor I}) 
+ S_2 ({\bi m}{\bi m}-{\bi \ell}{\bi \ell}) 
\en 
where ${\tensor I}= \{\delta_{ij}\}$ is the unit tensor. 
The orientation is uniaxial for $S_2=0$ 
but becomes  biaxial for $S_2\neq 0$  \cite{Gen93}.  
In terms of $S_1$ and $S_2$ in Eq.(2.17), 
$J_2$ and $J_3$ in Eq.(2.2) 
are calculated  as  
\be
J_2=\frac{2}{3}S_1^2+2S_2^2, \quad 
J_3= \frac{2}{9}S_1^3- 2S_1S_2^2.
\en
Here, if we  set  
\be 
S_1= S \cos\chi,\quad 
S_2= 3^{-1/2} S \sin\chi,  
\en 
 we   find 
\be 
J_2=\frac{2}{3}S^2,\quad J_3=  \frac{2}{9}
S^3 \cos(3\chi), 
\en
where the angle $\chi$ is in the range 
$-\pi/3 \leq \chi \leq \pi/3$  from $S_1\geq |S_2|$. 
As is well-known, when $F_0$ in Eq.(2.3) 
 is minimized with $B>0$,  the uniaxial orientation  
$\chi=0$ (or $S_2=0$) is selected and  $S(=S_1)$ is determined by 
\be 
{2C}S^2-{BS}-3A=0.
\en  
Since $S>0$ we have 
$
S= B/4C +[(B/4C)^2+ 3A/2C]^{1/2}.   
$

Far below the transition temperature, 
 the orientation is 
uniaxial and the amplitude $S$ 
 may be treated as a positive 
constant  outside the defect-core region. 
Mathematically, this limit can be conveniently achieved  
if we take the limit of large $A$ with 
$B/A$ and $C/A$  held fixed. The free energy 
$F_0+F_g$ is then approximated by 
the  Frank free energy,  
\be 
F_{\rm F}= \int d{\bi r}\frac{1}{2}K\sum_{ki}
 (\nabla_kn_i)^2,
\en 
with the single Frank coefficient,  
\be 
K= 2S^2L.
\en    
In this limit the dielectric tensor (2.13) is of the standard 
uniaxial form \cite{Gen93},  
\be 
\ve_{ij}
=\ve_\parallel n_in_j + 
\ve_\perp (\delta_{ij}-n_in_j), 
\en 
where   
$\ve_{\parallel}= \ve_0+ 2 \ve_1 S/3$  
and  
$\ve_{\perp}= \ve_0-  \ve_1 S/3$. 
From Eq.(2.15) an infinitesimal change of the director 
${\bi n}\rightarrow {\bi n}+\delta{\bi n}$ 
induces a change in  $F_e$ 
given by  
\be 
\delta F_e= -\int d{\bi r} \frac{\ve_1S}{4\pi}({\bi E}\cdot{\bi n}){\bi E} 
\cdot{\delta}{\bi n} .
\en 
Thus the equation for $\bi n$ is written as  \cite{NATO,Onu04}
\be 
K (\nabla^2{\bi n})_\perp 
+ \frac{\ve_1 S}{4\pi}({\bi E}\cdot{\bi n})
({\bi E})_\perp ={\bi 0}, 
\en 
where $(\cdots)_\perp$ denotes taking the 
perpendicular part of the vector ($\perp \bi n$).  
In the numerical 
analysis in this paper, however, 
we will  not use the above equation 
in terms of $\bi n$, 
 because the defect-core structure 
can  be better described in terms of $Q_{ij}$ 
 \cite{Scho,Hess}.

\subsection{Axisymmetric orientation}

As an application of our general framework, 
we focus on the 
basic problem of a single charged particle 
immersed in a nematic liquid crystal. 
The electric field around the particle  induces 
a  deformation of 
the orientation order parameter.
Far from the particle  
the orientation  is assumed to be uniaxially 
 along the z-axis. Before treating this 
case we here present the equations for 
$Q_{ij}$  for general  
 axisymmetric orientation.

It is convenient to use  the cylindrical 
coordinates $({\rho},\varphi,z)$, 
where $\rho=(x^2+y^2)^{1/2}$ 
and $\tan \varphi= y/x$.  Using 
 the unit vectors 
${\bi e}_\rho =(x/\rho,y/\rho,0)$, 
${\bi e}_\varphi= 
(-y/\rho,x/\rho,0)$, and ${\bi e}_z=(0,0,1)$,  
we express the traceless orientation tensor  as 
\bea 
\tensor{Q}&=&Q_1 {\bi e}_\rho {\bi e}_\rho 
+Q_2 {\bi e}_z{ \bi e}_z 
+Q_3( {\bi e}_\rho {\bi e}_z+ {\bi e}_z {\bi e}_\rho )
\nonumber\\
&-&(Q_1+Q_2) {\bi e}_\varphi {\bi e}_\varphi,  
\ena 
where 
$Q_i$   ($i=1,2,3)$  
depend on  $\rho$ and $z$.  
The eigenvalues of ${\tensor Q}$ 
are given by $\lambda_\pm= 
(Q_1+Q_2)/2\pm [Q_3^2+ (Q_1-Q_2)^2/4]^{1/2}$ 
and $\lambda_\varphi= -Q_1-Q_2$. 
Here we  limit ourselves to the case 
$\lambda_+ \geq \lambda_\varphi$ or 
\be 
{3}(Q_1+Q_2)+ [4Q_3^3+ (Q_1-Q_2)^2]^{1/2}\geq 0, 
\en 
under which $S_1$ and $S_2$ in Eq.(2.17) 
are written  as 
\begin{eqnarray}
S_1=\frac{3}{4} [4Q_{3}^2+(Q_{1}-Q_{2})^2]^{1/2}
+\frac{3}{4}(Q_{1}+Q_{2}),\\
S_2=\frac{1}{4}
[4Q_{3}^2+( (Q_{1}-Q_{2})^2]^{1/2}-\frac{3}{4}(Q_{1}+Q_{2}).
\label{eqI16}
\end{eqnarray}
With these definitions 
we shall find  $S_1\geq S_2\geq  0$  
everywhere in our numerical calculations,  
where the equality $S_1=S_2$ holds in an exceptional 
case  (see Eq.(4.10)).  
The director ${\bi n}$ is 
perpendicular to ${\bi e}_\varphi$ 
and the two components 
$n_\rho= {\bi e}_\rho\cdot{\bi n}$ 
and 
$n_z= {\bi e}_z\cdot{\bi n}$ satisfy 
\be 
\frac{n_\rho}{n_z} = 
\bigg \{ [(Q_{1}-Q_{2})^2+4Q_{3}^2]^{1/2}+Q_{1}-Q_{2}
\bigg \}\frac{1}{2 Q_{3}} .
\en
In terms of $S_1$, $S_2$, $n_\rho$, and $n_z$, 
we may express $Q_i$ as 
\bea 
Q_1 &=& (S_1+S_2)n_\rho^2 -\frac{1}{3}S_1-S_2, \\
Q_2 &=& (S_1+S_2)n_z^2 -\frac{1}{3}S_1-S_2,\\
Q_3 &=& (S_1+S_2)n_\rho n_z.
\ena  
In particular, $Q_1+Q_2= S_1/3-S_2$. 
In our numerical analysis to follow, 
 the condition (2.28) 
will be satisfied. However, 
if the reverse relation of Eq.(2.28) holds, 
we have $2S_1/3= -Q_1-Q_2>0$ and 
${\bi n}={\bi e}_\varphi$.

From Eq.(2.19)  $J_2$ and $J_3$ are written as   
\begin{eqnarray}
J_2&=&2(Q_1^2+Q_2^2+Q_3^2+Q_1Q_2),\nonumber \\
J_3&=&3 (Q_1+Q_2)(Q_3^2-Q_1Q_2) .
\end{eqnarray}
The  $F_0$  can then be expressed in terms of $Q_i$.  
The gradient free energy $F_g$ in Eq.(2.4) 
is of the form, 
\bea
F_g&=&L\int d{\bi r} 
\bigg[(\nabla Q_1)^2+(\nabla Q_2)^2+(\nabla Q_3)^2
\nonumber\\
&+&\nabla Q_1\cdot \nabla Q_2+ 
\frac{(2Q_1+Q_2)^2}{\rho^2}+\frac{Q_3^2}{\rho^2}\bigg], 
\ena 
where $(\nabla Q_i)^2= 
(\p Q_i/\p z)^2+ (\p Q_i/\p \rho)^2$ 
and the last two terms in the integrand 
$(\propto \rho^{-2}$) arise from 
the relations $\p {\bi e}_\rho/\p \varphi= {\bi e}_\varphi$ 
and  $\p {\bi e}_\varphi/\p \varphi= -{\bi e}_\rho$.

Also the  electric potential 
$\phi=\phi(\rho,z)$  is a function of $\rho$ and $z$ 
and the electric field is expressed as 
\be 
{\bi E}= E_1 {\bi e}_\rho + E_2{\bi e}_z, 
\en 
where $E_1=- \p \phi/\p \rho$ and $E_2=- \p \phi/\p z$.
The electric induction ${\bi D}= {\tensor{\ve}}\cdot{\bi E}$ 
 is expressed as 
\bea 
{\bi D}&=& 
[(\ve_0+\ve_1Q_1)E_1+ \ve_1Q_3E_2] {\bi e}_\rho \nonumber\\
&+& 
[(\ve_0+\ve_1Q_2)E_2+\ve_1Q_3E_1]{\bi e}_z.
\ena 
The Poisson equation (2.12) becomes 
\bea 
&&\hspace{-1cm}
\frac{1}{\rho}\frac{\p}{\p\rho}\rho
\bigg[(\ve_0+\ve_1Q_1)\frac{\p \phi}{\p\rho}+\ve_1Q_3\frac{\p\phi}{\p z}
\bigg ]\nonumber\\
&&\hspace{-1cm}+\frac{\p}{\p z}
\bigg[(\ve_0+\ve_1Q_2)\frac{\p \phi}{\p z}+\ve_1Q_3\frac{\p\phi}{\p \rho}
\bigg ]= -4\pi\hat{\rho}. 
\ena

We now set up the equilibrium 
equations for $Q_i$  requiring 
$\delta F/\delta Q_i=0$. 
From Eq.(2.15) we derive  the relation,   
\be 
\delta F_e =
 - \frac{ \ve_1}{8\pi}\int d{\bi r}
  ( E_1^2 \delta Q_1 +E_2^2
\delta Q_2 +2E_1E_2 \delta Q_3). 
\en 
Some calculations yield 
\bea 
&&\hspace{-12mm}\bigg({D_2} -L\nabla^2+\frac{4L}{\rho^2}\bigg ) 
(2Q_1+Q_2)-{D_3} = 
\frac{\ve_1}{8\pi } E_1^2, \\
&&\hspace{-12mm}\bigg (D_2-L\nabla^2+\frac{2L}{\rho^2}
\bigg ) (2Q_2+Q_1)- D_3 =
\frac{\ve_1}{8\pi}E_2^2, \\
&&\hspace{-12mm}\bigg (D_2-L\nabla^2+\frac{L}{\rho^2}\bigg )Q_3 = 
\frac{\ve_1}{8\pi}E_1E_2, 
\ena  
The left-hand sides are $\delta F_{\rm L}/\delta Q_i$ with 
$F_{\rm L}=F_0+F_g$ 
and the right-hand sides are $-\delta F_{e}/\delta Q_i$. 
We introduce 
\bea 
D_2&=&-A+CJ_2-B(Q_1+Q_2),\nonumber\\
D_3&=&B(2Q_1^2+2{Q_2}^2+5Q_1Q_2-{Q_3}^2).
\ena
Here we may define the amplitude 
$S= (3J_2/2)^{1/2}$  and the angle $\chi$ 
as in  Eq.(2.19). 
Then some calculations give 
\be 
{3\sqrt{3}}[4Q_3^3+ (Q_1-Q_2)^2]^{1/2} D_3= {2} B S^3 \sin (3\chi)
\en 
Thus $D_3 \propto (\p J_3/\p \chi)_S$, so 
$D_3=0$ leads to the uniaxial 
orientation $S_2=0$.  Further requirement of  
$D_2=0$ is to impose Eq.(2.21).

\section{Estimations of the free  energy} 
\setcounter{equation}{0}

\subsection{An isolated   charged particle}

We estimate the free energy contributions 
around an isolated  charged particle with charge $Ze$ and 
radius $R$ deeply in the nematic state.

If $\ve_1=0$, no orientation disturbance is induced and 
the electric potential is given by $Ze/\ve_0r$, where $r$ 
is the distance from the particle center. In this case, 
since ${\bi E}^2= (Ze/\ve_0)^2r^{-4}$,  
the electrostatic free energy $F_e$ 
is dependent on the lower cut-off radius $R$ as 
\be 
 F_{\rm B}=\int_{r>R} d{\bi r}\frac{\ve_0}{8\pi}{\bi E}^2 
= \frac{ Z^2e^2}{2\ve_0R}. 
\en 
This  is the first theoretical expression 
for the solvation free energy of ions 
in a polar fluid, where the lower bound 
$R$ is called the Born radius 
 \cite{Born,Is,Kit04}. For not large $\ve_1$ 
we may  expand  $ F_e$   as 
\be
 F_e =F_{\rm B}
- \int d{\bi r}
\sum_{ij}  
 \frac{ \ve_1}{8\pi} E_iE_j Q_{ij} +O(\ve_1^2), 
\en  
which follows from Eq.(2.15) at fixed charge density. 
For the case  $|\ve_1|\ls \ve_0$ 
we use the above expansion with 
${\bi E}= -(Ze/\ve_0r^2){\bi r}$  to estimate $F_e$. 
Further assuming that the orientation is 
nearly uniaxial, we obtain  
\be 
 F_e -F_{\rm B}\cong  -
 \ve_1S\frac{ Z^2e^2}{8\pi\ve_0^2} 
\int_{R<r<\ell} d{\bi r}\frac{1}{r^4} ({\bi n}\cdot{\hat{\bi r}})^2.  
\label{eq:20}
\en 
For $\ve_1>0$  (for $\ve_1<0$),  
$\bi n$ tends to be parallel 
(perpendicular) to 
 $\hat{\bi r}=r^{-1}{\bi r}$ 
near the charged particle  $R<r<\ell$, while 
$({\bi n}\cdot{\hat{\bi r}})^2$ should be replaced 
by the angle average  $1/3$  far from the particle 
$r>\ell$.   
We assume that the orientation disturbance 
is strong in the region $R<r<\ell$. On the other hand, 
the  Frank   free energy $ F_{\rm F}$ in 
Eq.(2.22) is roughly of order  
$ 2\pi K (\ell-R)$ \cite{estimate}. For $\ve_1>0$ 
 the change of the 
total free energy $F$ due to the orientation deformation 
 is estimated as  \cite{NATO,Onu04}
\be 
\Delta F  \cong \frac{2\ve_1}{3} \frac{S Z^2e^2}{2\ve_0^2}
\bigg (\frac{1}{\ell}-\frac{1}{R}\bigg ) +
2\pi K (\ell-R) . 
\en 
For $\ve_1<0$ 
the factor $2\ve_1/3$ of the first term should be replaced by 
$|\ve_1|/3$.  However,  
the  numerical  factors  of the two terms  in Eq.(3.4) 
are rough estimates and 
should not be taken too seriously.

Minimization of the first two terms  on the right-hand side 
of Eq.(3.4) with respect to $\ell$ 
 gives 
\be
\ell =
Ze (|\ve_1|S/{6\pi\ve_0^2K})^{1/2} . 
\en  
Strong orientation 
deformation occurs for 
$\ell>R$, which may be called the strong solvation 
condition for an isolated charged particle in 
liquid crystals.  
To characterize the strength of the charge 
we hereafter use  the  length, 
\be
\lambda_e=Z({\ell_B k_BT}/{8\pi K})^{1/2},  
\en 
where $\ell_B=e^2/\ve_0 kT$  is the Bjerrum length 
 usually of order of $10$ nm for liquid crystals. 
In terms of $\lambda_e$ 
 the strong solvation condition $\ell>R$ is written as 
\be 
(|\ve_1|/\ve_0)^{1/2}\lambda_e/R >1 
\en  
for $S \sim 1$. The left-hand side of Eq.(3.7) represents 
the dimensionless strength of the charge-induced deformation 
\cite{anchor}.
In  the reverse case 
$ (|\ve_1|/\ve_0)^{1/2}\lambda_e/R <1 $, 
the orientation deformation is weak, which may be called the 
weak solvation condition.

From Eq.(3.4) 
  the minimum 
(equilibrium) value of $\Delta F$ 
is  written as
\be 
\Delta F= 2\pi K( 2\ell- \ell^2/R -R),   
\en  
in terms of $\ell$ in Eq.(3.5). 
Here the right-hand side 
is zero for $\ell=R$ and decreases for larger $\ell$. 
Of course, even if Eq.(3.7) does not hold, 
weak deformations are induced to make 
$\Delta F<0$ as long as  $\ve_1\neq 0$. 
Notice that we are neglecting such weak deformations 
in the present estimations. 
See  Fig.14 below 
for numerical data of $\Delta F$.

\subsection{A  charged colloidal  particle}

Although it  is not clear whether or not 
ionization can be effectively 
induced on colloid  surfaces 
 in liquid crystals, 
we here assume the presence of charged colloidal 
paticles  in nematics. In such situations,  
the distortion of $\bi n$ 
due to the surface charge can be more important 
than that due to the anchoring 
 interaction given in Eq.(1.1).
The strong solvation condition  (3.7) 
is satisfied  for large $Z/R$, for example, when  
 the ionizable points on the surface 
is proportional to the surface area $4\pi R^2$. 
However,  the problem can be very complex,  
because the small counterions   
can  induce  large deformations  
of the orientation order   around themselves.

For simplicity, let us assume that the counterions do not  
satisfy Eq.(3.7) and only weakly disturb 
the orientation order  and   that the screening length 
$\lambda_s$  of  the colloid charge is  shorter than $R$. 
Then the distribution of the counterions is 
close to that near a planar charged surface and 
$\lambda_s$ is given by the Gouy-Chapman 
length  \cite{Netz},
\be 
\lambda_s = 2k_{ B}T/eE_{\rm s}= 1/2\pi \ell_{\rm B} \sigma_s, 
\en 
where $\sigma_s=Z/4\pi R^2$  is 
the surface charge density in units of $e$ 
and $E_{\rm s}= 4\pi \sigma_s e/\ve_0$ 
is the electric field  
at the surface.  To ensure  the inequality 
$\lambda_s <R$  we require 
\be 
2\pi \ell_{\rm B}R \sigma_s> 1.
\en  
Then $F_e-F_{\rm B}$ in  Eq.(3.3) may be written 
in the same form as $F_a$ in  
Eq.(1.1), 
\be 
F_a' = - \frac{1}{2} W_e \int dS
 ({\bi n}\cdot{\hat{\bi r}})^2.
\en 
Using Eq.(3.9) we obtain  \cite{NATO,Onu04}
\bea 
W_e &=& {S\ve_1Z^2e^2\lambda_s}/{4\pi\ve_0^2}R^4\nonumber\\
&=&  2k_BT S\sigma_s{\ve_1}/{\ve_0},
\ena 
which is independent of $R$ if $\sigma_s$ is 
a constant or $Z\propto R^2$. 
The  anchoring   due to the surface charge 
becomes  strong   for  $|W_e| R/K>1$  
with $K$ being the Frank constant 
in  Eq.(2.23). This is analogous to the well-known 
strong anchoring condition 
$|W_a| R/K>1$ for the neutral case 
\cite{Lav00,Ter95,Lub98,Yama,Sta04,Fuk04}.

\setcounter{equation}{0}
\section{Numerical calculations}

For  an isolated  spherical particle without the counterions, 
we   numerically solved 
Eqs.(2.41)-(2.43) satisfied by the three 
 components $Q_i$ and  the 
Poisson equation (2.39)  for the 
electric potential $\phi$, without the microscopic anchoring 
interaction in Eq.(1.1).   The particle radius is assumed to be 
considerably larger than the defect core radius.

\subsection{Method}

 To calculate  the equilibrium 
$Q_i$ and $\phi$, we 
integrated the  time-evolution equations,
\bea 
\frac{\p Q_i}{\p t}&=& - \frac{\delta F}{\delta Q_i}, 
\\ 
\frac{\p \phi}{\p t}&=& - \nabla\cdot{\bi D}+ 4\pi\hat{\rho}.   
\ena 
See  Eqs.(2.41)-(2.43) and 
the sentence below them for $\delta F/\delta Q_i$ 
and the left-hand side of 
Eq.(2.39) for $\nabla\cdot{\bi D}$.  
The steady solutions  reached at long times 
are the equilibrium or metastable solutions.  
In the following figures 
we will show the steady solutions only.

We used  a discretized $200\times 200$ cell 
in  the $(\rho,z)$ semi-plane 
($0\leq \rho\leq 200\Delta x$  and 
$0\leq z\leq 200\Delta x$) assuming 
the symmetry around the $z$ axis and 
with respect to the $xy$ plane (see the comment (i) 
in the last section). 
      In $F_{\rm L}$ 
we set $B/A=1$  and   $C/A=3$.  We took  the 
mesh size of the grid at 
\be 
\Delta x=(L/A)^{1/2}/2=\xi/2.  
\en 
We will measure space in 
units of $\Delta x$ in the following figures. 
The length $\xi= 
(L/A)^{1/2}$ is the shortest one 
in our theory and is of the order of 
the defect core size. 
The particle radius $R$ was fixed as  a relatively large value, 
\be 
R= 10\xi= 20\Delta x.
\en  
All the figures to follow  will be given in the region 
$-100\leq x\leq 100$  and $-100\leq z\leq 100$ 
with the $x$ and  $z$ axes  being horizontal and vertical,  
respectively.

On all the cell boundaries, we assumed the constant-potential 
condition $\phi=0$ in Eq.(2.7) 
and the uniaxial orientation along the $z$ axis,  
\be 
Q_1= -\frac{1}{3} S_{\rm b},\quad 
Q_2= \frac{2}{3} S_{\rm b},\quad
Q_3=0.
\en 
Here $S_{\rm b}\cong 0.79$ 
is the bulk amplitude obtained as the solution of Eq.(2.21). 
We  calculated  $Q_i$ only  outside the sphere $r>R$. 
If we superimpose  $\delta Q_{ij}$ on $Q_{ij}$ with 
$\delta Q_{ij}=0$ on the system boundaries,  
the incremental change of $F$ is written as 
\be 
\delta F= -L\int da \sum_{ij} ({\bi \nu}
\cdot \nabla Q_{ij})\delta Q_{ij}, 
\en 
where bulk contribution vanishes,  
$da$ is the surface element,  and 
 ${\bi \nu}$ is the  outward normal unit vector 
on the particle surface (equal to 
$r^{-1}{\bi r}$ here).  
To ensure $\delta F=0$ we thus imposed 
\begin{equation}
{\bi \nu}\cdot 
\nabla Q_{ij}  =0,  
\end{equation} 
at $r=R$.   This becomes 
${\bi r}\cdot\nabla Q_i=\rho \p Q_i/\p\rho+ z\p Q_i/\p z=0$ 
at $r=R$  in the axisymmetric case in Eq.(2.27). 
We   assume  no  surface free energy 
or  no supplementary 
anchoring on the particle surface. That is, 
 we set $W_a=0$ in Eq.(1.1). 
On the other hand,  the electric 
potential $\phi$  was calculated 
in the whole region in the cell  for the computational  
convenience. That is,  we assumed isotropic  polarizability  
(${\bi D}=\ve_0{\bi E}$)  
inside the sphere $r<R$ 
and used  the smooth 
charge-density profile,
\be
\hat{\rho}({\bi r}) =\frac{1}{2}\hat{\rho}_0-
\frac{1}{2}\hat{\rho}_0 \tanh[(r-R)/\xi],
\en  
in the whole region. 
The constant $\hat{\rho}_0$ 
is determined from $\int d{\bi r}\hat{\rho}({\bi r})= Ze$.  
This means that  $\phi$ obeys 
$-\ve_0\nabla^2\phi=4\pi\hat{\rho}$ 
for $r<R$  and  Eq.(2.39) for $r>R$.  
As the steady solutions of Eq.(4.2), 
we confirm that 
$\phi$ and ${\bi r}\cdot{\bi D}$ 
change continuously across the interface 
 even on the scale of the mesh size $\Delta x$, 
which are required in electrostatics.

Since $R/\xi$ is fixed at 10,  
the  remaining 
 relevant  control parameters  are the ratios 
 $\lambda_e/R$ and $\ve_1/\ve_0$. 
We thus  varied $\lambda_e$ at $\ve_1/\ve_0=\pm 1$.

\begin{figure}[h]
\includegraphics[scale=0.4]{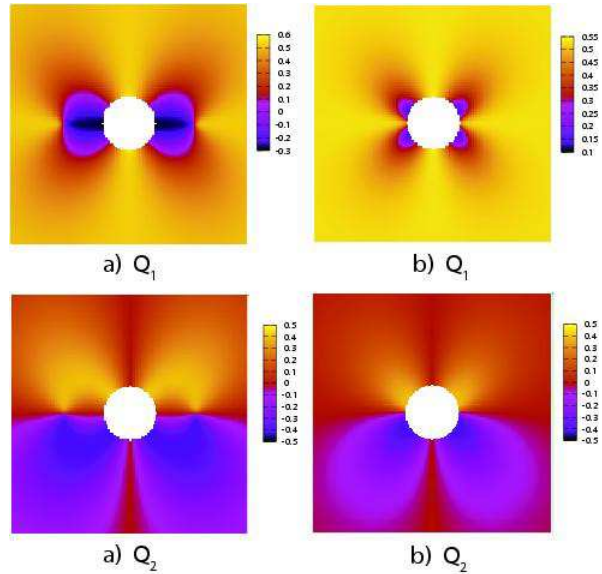}
\caption{ 
$Q_{2}$ and $Q_3$ for  (a) 
$\lambda_e/R=10$  and (b) $5$  at $\ve_1/\ve_0 =1$  in the range 
$-100\leq  x\leq 100$ and $-100\leq z\leq 100$ in units of 
$\xi/2$ (see Eq.(3.2)). The $z$ axis is vertical and the 
$x$ axis is horizontal.  There is a Saturn-ring defect in (a) 
and no defect in (b). 
}
\end{figure}

\subsection{Orientation  for $\ve_1>0$}

\begin{figure}[bt]
\includegraphics[scale=0.35]{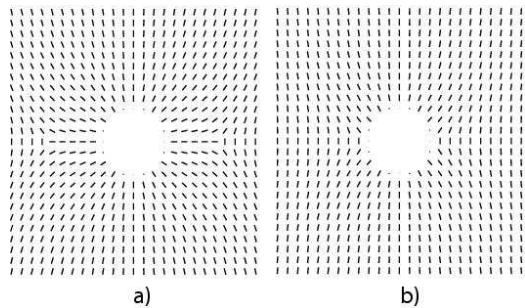}
\caption{
Nematic director $\bi n$ in the $xz$ plane 
 for (a) $\lambda_e/R=10$  and (b) $5$  
at $\ve_1/\ve_0=1$. A Saturn ring 
is present  in (a) and is nonexistent in (b).}
\end{figure}

\begin{figure}[t]
\includegraphics[scale=0.4]{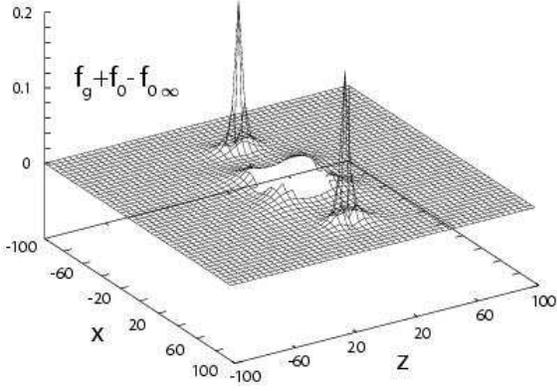}
\caption{
Landau-de Gennes free energy density 
$f_g+f_0-f_{0\infty}$   for 
$\lambda_e/R=10$ and $\ve_1/\ve_0=1$  in the 
$xz$ plane, where $f_{0\infty}$ 
is the background value of $f_0$.  It 
is peaked at the Saturn ring.}
\end{figure}

\begin{figure}[h]
\includegraphics[scale=0.4]{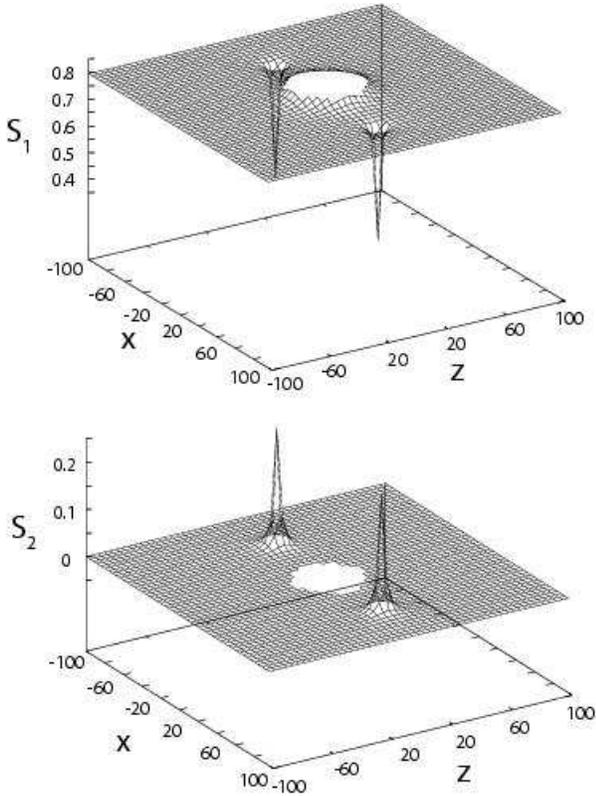}
\caption{
Scalar order parameter $S_1$   
and biaxiality parameter $S_{2}$ for $\lambda_e/R=10$ 
and  $\ve_1/\ve_0=1$   in the $xz$ plane. 
}
\end{figure}

We first focus on the case $\ve_1/\ve_0=1$, where 
the parallel alignment 
$\bi n \parallel {\bi E}$  
is favored near the particle. 
This corresponds to the homeotropic alignment 
(perpendicular to the surface) in the 
neutral  case. 
In Fig.1,  we  display  
$Q_{2}$ and $Q_{3}$ in Eq.(2.27)  in the $xz$ plane  for   
(a) $\lambda_e/R=5$  and  (b) $10$. 
These quantities are related to 
$\bi n$ as in Eqs.(2.32) and (2.33). 
  Figure 2
 displays the corresponding configuration of ${\bi n}= 
n_\rho{\bi e}_\rho+ n_z{\bi e}_z$ in the $xz$ plane 
deduced from Eq.(2.31). For  large  $\lambda_e/R$,
the nematic exhibits a radial orientation at the surface
of the particle and 
a line of $-1/2$ defect  appears  surrounding 
 the particle  in the plane  $z=0$,  because the orientation 
is along the $z$ axis far from the particle. 
Such a circular defect line is  called "Saturn ring" 
in the literature \cite{Lav00,Ter95,Lub98}. 
In Fig. 2,  we  can see  a -1/2 defect    on both  sides  of 
the particle on the $x$ axis  
for $\lambda_e/R=10$ in (a), but there is no defect 
for  $\lambda_e/R=5$ in (b).

In Fig.3,  we  plot  the deviation of the Landau-de Gennes 
free energy density $f_g+f_0-f_{0\infty}$  at $\lambda_e/R=10$, 
 where $F_0=\int d{\bi r}f_0$ and  $F_g=\int d{\bi r}f_g$ 
are given in Eqs.(2.3) and (2.4).  The  $f_{0\infty}$  
is the value of $f_0$ far from the particle in the 
uniaxial state  with  $S_1=S_b$ and $S_2=0$ (see Eq.(4.1)). 
This figure  clearly demonstrates   the 
existence of a  Saturn ring at the sharp peaks.    
Next, in Fig.4, we display  $S_1$ in Eq.(2.29) and $S_2$ in Eq.(2.30), 
respectively,  for  $\lambda_e/R=10$. At the defect positions, 
 $S_1$ becomes small and $S_2$ exhibits a  peak, 
while they tend to  their bulk values, $S_1 = 0.79$  and $S_2= 0$, 
far  from  the defect.  In the defect region,  
the nematic is locally melt  and   biaxial. 
The latter biaxial  behavior was also found 
in the previous molecular dynamics 
simulation for a neutral particle under the 
homeotropic anchoring  \cite {Lav00,Allen}.

We performed simulations for  various other 
parameters (not shown 
here).  With  increasing $\lambda_e/R$, a Saturn ring 
appears suddenly with a nonvanishing radius 
at  a certain  transition 
value of $\lambda_e/R$ (see Fig.9 below). 
Essentially the same  behavior 
can be observed  with increasing  
$\ve_1/\ve_0$ at fixed $\lambda_e/R$.
At small  $\lambda_e/R$, 
 the coupling between the electric field and 
the  nematic order is not strong enough to 
induce the radial orientation on the equatorial
plane. We therefore observe the  
quadrupolar symmetry without
defect formation  for $\lambda_e/R=5$  in Figs.1 and 2, 
where  $S_1$, $S_2$ and $f_0+f_g$ 
 exhibit  no peaks around the particle.

\begin{figure}[h]
\includegraphics[scale=0.4]{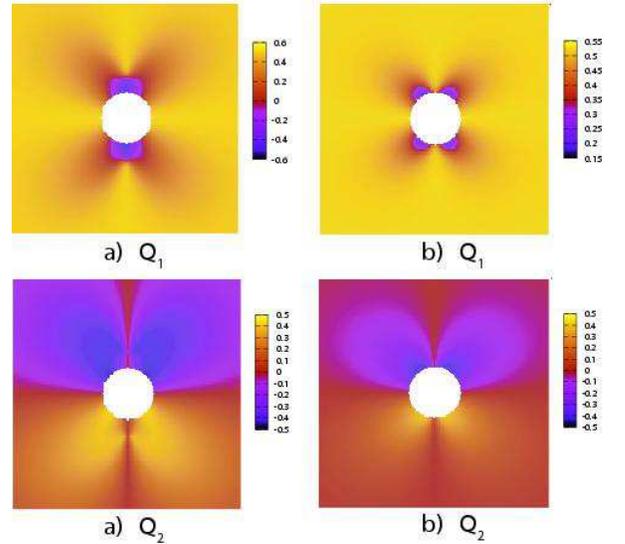}
\caption{
$Q_{2}$   and $Q_3$ for 
(a)  $\lambda_e/R=10$  and (b) $4.2$  
at $\ve_1/\ve_0=-1$  in the $xz$ plane.
}
\end{figure}

\begin{figure}[h]
\includegraphics[scale=0.35]{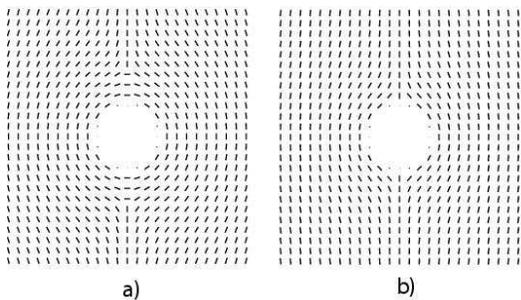}
\caption{
Nematic director $\bi n$ 
 for (a) $\lambda_e/R=10$  and (b) $4.2$  
at $\ve_1/\ve_0=-1$  in the $xz$ plane. 
A pair of point defects can be seen in (a), 
while there is no defect in (b).}
\end{figure}

\begin{figure}[h]
\includegraphics[scale=0.4]{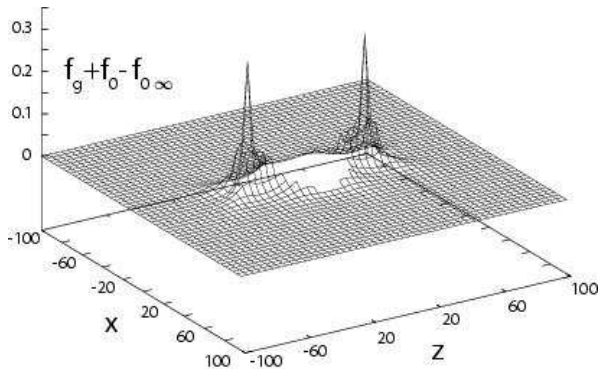}
\caption{ 
Landau-de Gennes free energy density 
$f_0+f_g-f_{0\infty}$ for 
$\lambda_e/R=10$ and $\ve_1/\ve_0=-1$  
in the $xz$ plane, where $f_{0\infty}$  is the 
background value of $f_0$.  
It is peaked along the disclination 
line segments connecting the poles and the point 
defects.
}
\end{figure}

\begin{figure}[h]
\includegraphics[scale=0.4]{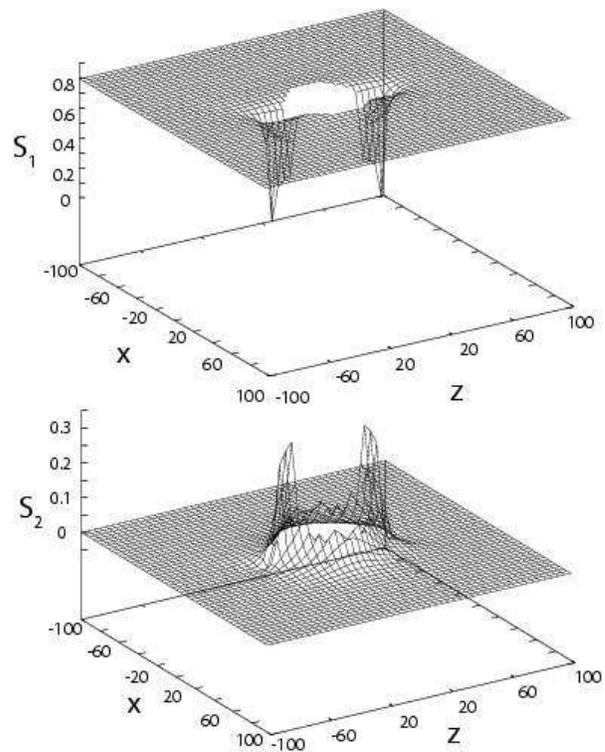}
\caption{
Scalar order parameter $S_1$ 
and biaxiality parameter $S_{2}$ 
for $\lambda_e/R=10$ and 
$\ve_1/\ve_0=-1$  in the $xz$ plane.  
They deviate from the background values 
around the disclination  line segments.
}
\end{figure}

\subsection{Orientation   for $\ve_1<0$}

We turn  to the second  case   $\ve_1<0$ 
at  $\ve_1/\ve_0=-1$, where    
the perpendicular alignment $\bi n \perp {\bi E}$ is preferred   
close to the particle. We  show   
 $Q_2$  and  $Q_3$  in Fig.5 
 and $\bi n$  in Fig.6   
for (a)  $\lambda_e/R=10$  and (b) $4.2$  
 in the $xz$  plane. Remarkably, in (a) the 
director is everwhere parallel to the surface 
and   a pair of point defects    are created 
on the $z$ axis, while in (b) 
the director is not parallel 
to the surface in the neighborhood 
of the poles.  With these defects in (a),  
 the Landau-de Gennes free energy density 
$f_0+f_g$  exhibits twin maxima  
 at the point-defects 
 as in Fig.7, while    $S_1$ and $S_2$ 
behave as in Fig.8.

The defect structure in (a) 
has never been reported in the literature. 
Each point defect is detached  from the surface 
and is connected to one of the poles by a  
$+1$ disclination line segment 
with length $\ell_d$.  Here $\ell_d=37$ in  
 (a).    In fact, on both sides, 
a ridge structure  
with its top at the point-defect  
can be seen in $f_0+f_g$, $S_1$, 
and $S_2$ in  Figs. 8 and 9. 
The singular  line segments   created are 
specified by $R<|z|<R+\ell_d$ and $\rho=0$, 
on which  our numerical analysis 
yields  $Q_3=0$  and   $Q_1=-Q_2/2>0$,   so that 
\be 
{\tensor Q}= Q_1 ( {\tensor I}- 
3{\bi e}_z{\bi e}_z  ). 
\en 
Here ${\tensor I}$ is the unit tensor. 
This form  is natural 
from the rotational invariance  around the $z$ axis 
and $n_z=0$.    On  these  line segments,  the equality, 
\be 
S_1=S_2=3Q_1/2, 
\en 
can be found from Eqs.(2.29) and (2.30). 
To see the defect structure 
in more dtail, we plot 
 $S_1$ and $S_2$ as functions of $z$  
for $\rho=0$ (along the $z$ axis) and $\rho=3$  in Fig.9  
and   $S_1$,  $S_2$, and $n_z$  as functions of $\rho$  
for $z=30$ in Fig.10. We can see how  $S_1$ becomes small 
and equal to $S_2$ along the $z$ axis, while $S_1>S_2$ 
not on the $z$ axis.   
We also recognize that $n_z$ changes slowly 
on the scale of $R$, while   $S_1$ and $S_2$ 
change    on the scale of the core radius.   A high degree 
of biaxiality is present around the line segments. 
In particular,   $S_1=S_2$ 
on  the singular line segments and 
$S_1>S_2$   outside them.

In the neutral  case with short-range anchoring, 
 a pair of  point defects, called 
boojums \cite{P1,Volovik}, are attached to 
the surface  at  the poles  for large negative $W_a$.

\begin{figure}[t]
\includegraphics[scale=0.4]{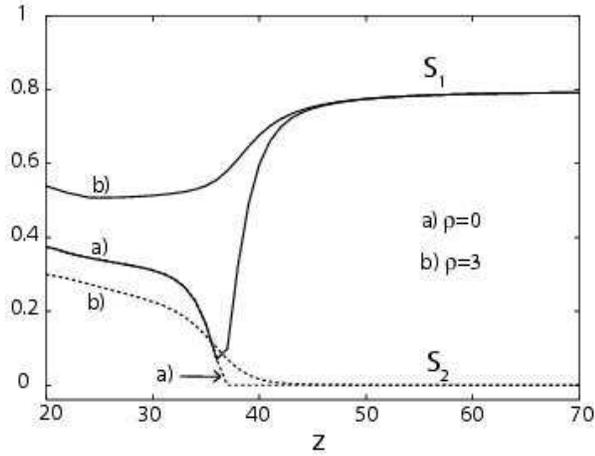}
\caption{$S_1$ (solid line) and $S_2$ (dotted line) 
along the $z$ axis   with  $\lambda_e/R=10$ and 
$\ve_1/\ve_0=-1$  for 
(a) $\rho=0$ and (b) $\rho=3$, where 
$S_1=S_2$  on  the line  segment 
$R<z<R+\ell_d$ at $\rho=0$. 
}
\end{figure}

\begin{figure}[th]
\includegraphics[scale=0.4]{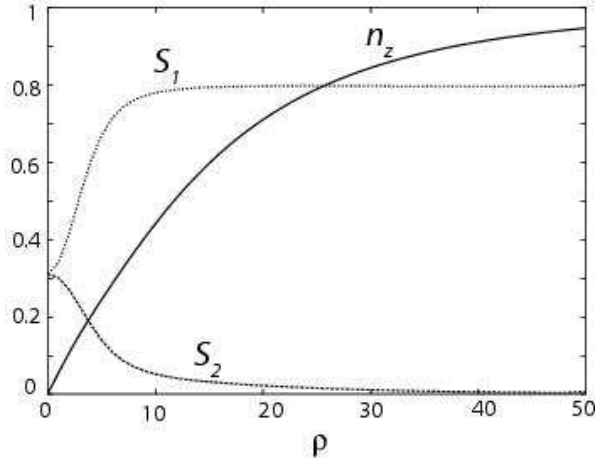}
\caption{$S_1$ (dotted line),  $S_2$ (dotted line), and $n_z$ (solid line) at $z=30$ along the $\rho$ axis  for $\lambda_e/R=10$  and $\ve_1/\ve_0=-1$.
}
\end{figure}

\subsection{Defect - no defect transition}

\begin{figure}[ht]
\includegraphics[scale=0.37]{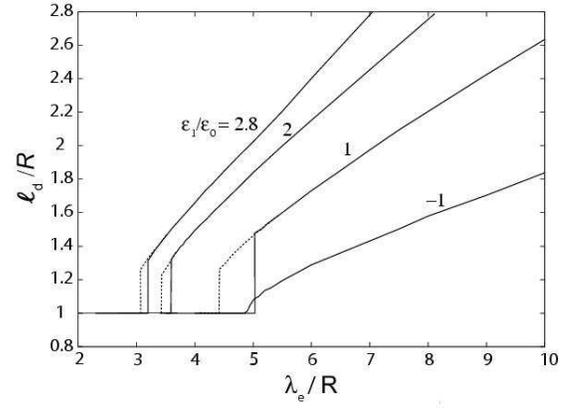}
\caption{Normalized defect radius $\ell_d/R$ 
vs  $\lambda_e/R$  for various $\ve_1/\ve_0$. 
For positive $\ve_1$  the defect formation  and collapse are  
discontinuous, where the solid and dotted curves 
are the results of  increasing and decreasing 
$\lambda_e$, respectively. For negative $\ve_1$ 
the transition  is  continuous with  a critical point.}
\end{figure}

\begin{figure}[t]
\includegraphics[scale=0.37]{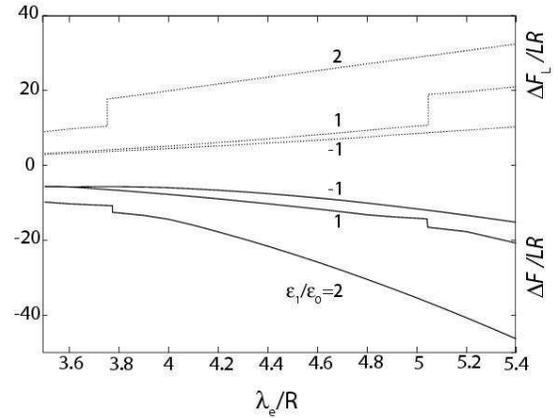}
\caption{Normalized changes of 
the total free energy Eq.(4.7) (solid lines) 
and the Landau-de Gennes free energy Eq.(4.6) 
(broken lines)  
vs $\lambda_e/R$  for three $\ve_1/\ve_0$. 
For $\ve_1>0$ the discontinuities correspond to the defect formation 
in Fig.11. 
}
\end{figure}

To gain  quantitative information,  
we  computed the distance of  
the defect core  from the particle center $\ell_d$ 
as a function of  $\lambda_e/R$ for various values of 
$\ve_1/\ve_0$.  As  shown in Fig.11,  a Saturn ring appears 
discontinuously  for positive 
$\ve_1$.  The solid curves 
were obtained when  $\lambda_e$  was increased 
from small values, 
while the  dotted  curves 
were obtained when  $\lambda_e$  was 
decreased from large values.  We can see that the  
value of $\lambda_e/R$  at the 
formation  is larger than that at the disappearance. 
Similar hysteresis behavior 
was also found in a two-dimensional simulation 
with positive $W_a$ by Yamamoto\cite{Yama}. 
For negative $\ve_1$, a pair of 
 defects appear from the  
surface continuously and is detached for 
$\lambda_e/R$  larger than the critical value
($\sim 4.9$ in Fig.11). For both positive and negative 
$\ve_1$, the transition value of $\lambda_e/R$ 
decreases with increasing 
$|\ve_1|/\ve_0$ (not shown  for negative $\ve_1$).  
These transitions are  
consistent with the criterion (3.7) 
since the defect formation is 
 possible only in the strong solvation condition. 
Furthermore, once the defect  is created, 
the defect size 
$\ell_d$ grows linearly with $\lambda_e/R$, 
with the slope increasing  with increasing 
$|\ve_1|/\ve_0$ (which 
is the case also for negative $\ve_1$).  
This trend is also consistent with 
Eq.(3.7), provided that 
 the distance of the orientation deformation 
$\ell$ there and the defect size $\ell_d$ here are 
assumed to be of the same order.

Next, in Fig. 12, we plot the   increase of 
the Landau-de Gennes free energy,  
\be 
\Delta F_{\rm L}=\int _{r>R}d{\bi r} 
[f_0+ f_g-f_{0\infty}], 
\en 
and that of the total free energy,  
\be 
\Delta F=F- F_{\rm un}
=\Delta F_L+ \Delta F_e, 
\en  
as a function of $\lambda_e/R$ 
for three values of $\ve_1/\ve_0$, where $\Delta F_e$ is 
the increase of the electrostatic part. 
 These quantities 
were  calculated  outside the particle $r>R$ 
using the solutions  of 
Eq.(2.39) and Eqs.(2.41)-(2.43).  The 
 $F_{\rm un}$ is the  value of $F$ 
in the unperturbed state in which 
the  orientation is uniaxially along 
the $z$ axis and the electric potential $\phi$ 
is calculated using the homogeneous 
$\ve_{ij}$ in the uniaxial state. Thus these increases  
 vanish as $\lambda_e/R \rightarrow 0$. 
We recognize that  $\Delta F_{\rm L}$  increases 
but $\Delta F$ decreases   
with increasing $\lambda_e/R$. These aspects are consistent 
with the estimations in Eqs.(3.4) and (3.8), 
though they are rough approximations.  
It is remarkable that $\Delta F_e$ is negative 
and its absolute value is larger than 
$\Delta F_{\rm L}$. 
In addition, at the Saturn-ring formation, 
$\Delta F/LR$  
changes  by $-1.1$ and $-0.9$ 
for $\ve_1/\ve_0=1$ and 2, respectively, 
while the corresponding changes of $\Delta F_{\rm L}/LR$  
are 8.2 and 7.1, respectively. 
Notice also that $\Delta F_{\rm L}$ and 
$\Delta F$  are continuous even at the defect formation 
for  negative $\ve_1$.

\section{Summary and discussions}

We summarize our main results. 
In Sec.2, we have  derived 
the equilibrium equations for the tensor order 
parameter $Q_{ij}$ as in Eq.(2.16),  supplemented with the Poisson 
equation (2.12).   The  interaction between the orientation and 
the electric field arises from the 
 dielectric anisotropy $\ve_1$ in Eq.(2.13).  
In the axisymmetric case $Q_{ij}$ is expressed in terms of 
the three components $Q_i$  ($i=1,2,3$) as in Eq.(2.27). 
They are determined by   Eqs.(2.41)-(2.43) 
 with the aid of the Poisson equation Eq.(2.39). 
In Sec.3, we have estimated the range $\ell$ 
of the strong orientation deformation 
around an isolated charged particle 
as in Eq.(3.5) and obtained the criterion of the strong 
deformation as in Eq.(3.7). There, we  can 
see analogy between the present problem of liquid crystals 
and that of the ion solvation 
in polar fluids \cite{Kit04}.   
For a highly charged colloidal particle, 
we have found the effective anchoring parameter  
Eq.(3.12) under the condition   Eq.(3.9). 
 In Sec.4, 
we have numerically examined the orientation deformation 
around an isolated charged particle. 
For positive  $\ve_1$  and for  large $\lambda_e/R\propto Z/R$, 
we have obtained  a Saturn ring around the particle, 
where $\lambda_e$ is the characteristic length in Eq.(3.6).
As shown in Fig.11,  the defect radius 
exhibits a discontinuous  
change as a function of $\lambda_e/R$ 
for positive  $\ve_1$.  This  means that a Saturn ring 
appears or disappears suddenly with radius larger than $R$  as 
$\lambda_e/R$ is increased  or decreased.  
For negative $\ve_1$ and  large $\lambda_e/R$,  
 we have found   appearance of point  defects (boojums) 
 on both sides of the particle  along the $z$ axis 
as a continous transition. 
They  are detached  from the  surface, 
while they are on the surface for a  neutral particle.
As shown in Fig.12, 
the  decrease of  the electrostatic free 
energy  overcomes the increase of the Landau-de Gennes 
free energy, resulting in the  lowering 
of the total free energy, with the orientation deformation. 
\\

We make further comments on  the limitations of 
our work and future problems.\\ 
(i) In our simulations we have sought  
only the axisymmetric defrmations 
symmetric with respect to the $xy$ plane. 
However, as in the neutral case \cite{Lav00,Ter95,Lub98,P1}, 
 there can be symmetry-breaking defects 
such as a combination of a radial 
and a hyperbolic hedgedog.\\
(ii) The orientation deformation occurs  
much longer than  the  particle radius. 
For microscopic ions the deformation 
extends over  a nanometer scale. 
From Fig.12 we can see that the lowering of the 
 free energy much exceeds  $LR  \sim k_BT R/a$,  
where $a$ is the size of a liquid crystal 
molecule.  Thus, even for $R \sim a$, we find 
 $|\Delta F| \gg k_BT$ and the deformation 
can be  stable against thermal fluctuations.   
As mentioned in Sec.1, 
the long-range deformation should be the 
 origin of  anomalously low ion mobility 
in nematics \cite{Gen93,PG}. 
However, the existence of defects on a nanometer scale 
is not well established because of the limitation 
of  our coarse-grained  approach. 
\\ 
(iii) We have examined the charge effect  only in 
nematics.  Slightly above the (weakly first-order)
isotropic-nematic phase transition, 
the deformation around charged particles 
can   be long-ranged, extending over the correlation 
length \cite{Fukuda}, as in  near-critical polar 
fluids \cite{Kit04}. 
Furthermore,  metastability 
or hysteresis  behavior at the isotropic-nematic transition 
should almost disappear 
 in the presence of a small amount of ions, 
owing to ion-induced nucleation, as in   polar fluids \cite{Kit05}.\\
(iv) We have presented numerical analysis 
only for a single particle, but 
long-range correlations among charged inclusions can produce 
a number of unexplored effects. 
In particular, doping of a small amount of ions into liquid 
crystals should provide 
 highly correlated ionic systems, where 
transparent nematic states would be realized \cite{Yam01}. 
 Ion solubility and ion distribution across 
 an isotropic-nematic interface 
should  constitute  new problems, which have been studied 
for  polar fluids \cite{Onuki2006}.\\
(v)  We should  investigate  dynamical  properties 
of charged particles in liquid crystals such 
 as the ion mobility or convection \cite{Gen93,Fukuda}. 
They  should be much influenced by the orientation 
deformation \cite{omoto}.   
\\ 
(vi) Beyond the particular interest 
in the field of liquid crystals, 
the general approach developed in this paper 
and preceding ones \cite{Kit04,Kit05,NATO,Onu04,Onuki2006} 
using inhomogeneous 
dielectric constant or tensor  
 could provide a practical and coherent method to 
study various polarization effects 
in simple and complex fluids.

\begin{acknowledgements}  

We  thank T. Omoto for  providing us his Master thesis 
on dynamics of ions in nematics \cite{omoto}. 
Thanks are also due to  J.-i. Fukuda, A. Furukawa, 
A. Minami,  
T. Nagaya, H. Tanaka, and R. Yamamoto 
  for valuable discussions.  
One of the authors (L.F.)  received 
financial support   from  JSPS  
during a  postdoctoral stay at  Kyoto University. 
This work was supported by Grants-in-Aid 
for scientific research and 
 the 21st Century COE project 
 from the Ministry of Education, 
Culture, Sports, Science and Technology of Japan.

\end{acknowledgements}


\begin{references}
\bibitem{anchor}   In the neutral case, 
the dimensionless anchoring strength 
is given by $W_aR/K$,  
where $K$ is the Frank coefficient in Eq.(2.22). 
Defects are formed for $|W_a|R/K \gs 10$.
\bibitem{nonlinear} In the vicinity of  
small charged particles,  
the electric field can be  very strong 
and the effect of nonlinear dielectric 
saturation needs to be considered.
\bibitem{estimate}  
Let us  set ${\bi n}=r^{-1}{\bi r}$ 
in the region $R<r<\ell$. Then the 
space integration of the Franck free energy density 
in Eq.(2.22) in this region becomes $2\pi(\ell-R)$.
\end{references}
\end{document}